\pdfoutput=1 
\documentclass[preprint,12pt]{elsarticle}

\usepackage{amssymb}
\usepackage{CJKutf8}

\journal{arXiv}
\newcommand{\comm}[1]{}

\begin{document}

\begin{frontmatter}



\title{Occupational Income Inequality of Thailand: A Case Study of Exploratory Data Analysis beyond Gini Coefficient}


\author[inst1]{Wanetha Sudswong}
\author[inst1]{Anon Plangprasopchok}
\author[inst1]{Chainarong Amornbunchornvej\fnref{label2}}

\affiliation[inst1]{organization={National Electronics and Computer Technology Center (NECTEC)},%Department and Organization
            addressline={112 Phahonyothin Road, Khlong Nueng}, 
            city={Pathum Thani},
            postcode={12120},
            country={Thailand}}

 \fntext[label2]{Corresponding author, email: chainarong.amo@nectec.or.th}

\begin{abstract}
% Summary - Gini is insufficient. EDOIF can help with this insufficiency.
Income inequality is an important issue that has to be solved in order to make progress in our society. The study of income inequality is well received through the Gini coefficient, which is used to measure degrees of inequality in general. While this method is effective in several aspects, the Gini coefficient alone inevitably overlooks minority subpopulations (e.g. occupations) which results in missing undetected patterns of inequality in minority.

%, an index which measures income inequality of an economy by observing the relationship between the cumulative population and the cumulative income of an economy with respect to the ideal standard of equality. 

In this study, the surveys of incomes and occupations from more than 12 millions households across Thailand have been analyzed by using  both Gini coefficient and network densities of income domination networks to get insight regarding the degrees of general and occupational income inequality issues.  The results show that, in agricultural provinces, there are  less issues in both types of inequality  (low Gini coefficients and network densities), while some non-agricultural provinces face an issue of occupational income inequality (high network densities) without any symptom of general income inequality (low Gini coefficients). Moreover, the results also illustrate the gaps of income inequality using estimation statistics, which not only support whether income inequality exists, but that we are also able to tell the magnitudes of income gaps among occupations. These results cannot be obtained via Gini coefficients alone. This work serves as a use case of analyzing income inequality from both general population and subpopulations perspectives that can be utilized in studies of other countries.

\end{abstract}


% \begin{highlights}
% \end{highlights}


\end{frontmatter}


\section{Introduction}
\label{sec:introduction}

% 1) - Important of income inequality
Income inequality exists in nearly every aspect of modern human lives from the perspective of individuals within an economy. It is an indicator of national economic development \citep{Dararatt2019}, integrated within the UN Sustainable Development Goals on Poverty, Decent Work and Economic Growth, and Reduced Inequalities \citep{SDG}. The alleviation of income inequality is the primary prerequisite for meeting \textit{basic human needs} on a global scale \citep{Jolly1976, Streeten1978, Moon1985}. Thus, income inequality is a global issue concerning all and is defined as the disproportionate income distribution among households \citep{Todaro2015}. 

%2) - Solution to income inequality -> human cap theory -> measure -> introduction of gini coef
 In \textit{Human capital theory}, it assumes that increasing the productive capacity of individuals can be done through education. The implication of the theory is that effective education and training can increase overall productivity and value of an individual \citep{Nafukho2004, Becker1993, Mincer1958}. However, before dealing with income inequality issues in a particular area, policy makers have to find how severe the issue is in the area. One of the well-known measures of income equality is the \textit{Gini coefficient}, a value representing adjacency to perfect income equality on a scale of 0, perfect equality, to 1, complete inequality \citep{IncomeInequalityMeasures, Xu2018, Dararatt2019}. A lower Gini coefficient indicates less impact on income inequality while a higher Gini coefficient is indicative of higher levels of inequality. With the light of human capital theory, policy makers can place a policy to improve skills of people in areas with high Gini coefficients in order to combat income inequality.

%3) - Introducing Thailand (leading with 3 support reasons) & share GDP sectors in the country -> Thailand is suitable case study
 Developing countries tend to display lower levels of human capital as well as higher levels of inequality and poverty \citep{Todaro2015}. Thailand is one of developing countries that is still encountering income inequality issue. Inequality in Thailand has been sporadically rising after 1992 \citep{Phonpaichit2016}. Studies point to three supporting issues: the top income households continue to increase gains in the form of income and property, the bottom income households struggle to develop human capital, and quality of formal education and training is below standards despite its increase in quantity \citep{Booth2019, Meneejuk2016, Wasi2019}. According to the Credit Suisse Global Wealth Databook for 2020, Thailand has been experiencing a severe decline in GDP since 2019 and is forecasted to continue its decline \citep{Suisse2020databook, nesdcGDP2021}. The share of GDP by industry in 2008 for Thailand shows that only 12\% goes to the agriculture sector, 44\% to the industrial sector, and the remaining 44\% to the services sector \citep{Todaro2015}. Over the years, this distribution varied and the GDP share of agricultural economic activities alone now comprise of only 8.4\% in 2017, 8.2\% in 2018, and 8.1\% in 2019 \citep{nesdcNI2021}. All proxies support the existence of general income inequality and disproportional income distribution by varying industries and occupation in Thailand. Therefore, Thailand is a suitable case study for the subject of income inequality and occupational income distribution.

%4) - Issue of Gini -> insight policy makers need & gini can't give -> occupation income inequality
Thailand shows indication of minimal human capital income development and suggested poverty via disproportional income distribution by occupation. However, the Gini coefficient alone reveals opposing results. The Gini coefficient of Thailand according to the National Statistics Office of Thailand is 0.421 in 1998, 0.444 in 1999, 0.439 from 2000, 0.419 in 2001, and 0.428 by 2002~\cite{NSOcoreEco}. Other studies support the finding that specify a decrease in income inequality between 2002 and 2015 with a decline from 0.508 to 0.445 \citep{Dararatt2019, Phonpaichit2016}. The net decline in values over time suggests a flourished society with lessening income inequality. This is inconsistent with GDP shares by industry where majority and minority groups are distinguished. Thus, the evidence implies a practicality for an additional measurement of income inequality along with the Gini coefficient. Additionally, to utilize human capital theory, policy makers have to know which occupations in areas have severe gaps of incomes compared to other occupations so that they can re-skill or improve education of occupations that face problems. Nevertheless, the Gini coefficient alone is unable to infer income gaps among different occupations. Provinces where the Gini coefficient are low indicate a well distributed population overall, however, Gini coefficients fail to recognize underlying forms of income inequality such as occupation-centered income inequality among the same population if the majority has no problem while some minority occupations face the issue of income inequality.

% 5) - explain general and occupational income inequality
In this paper, we propose two distinguishable types of income inequality: general income inequality of a population and occupational income inequality within the same population. Both are indicative of the distribution of wages across a population. However, only the newly proposed occupational income inequality takes into account the majority and minority of the population. The categorization of these groups within a population is separated by occupations declared in Section \ref{OccuCategory}.\textit{General income inequality}, used interchangeably with income inequality, refers to the overall measure of economical distribution of a population. The measure of this is the Gini coefficient. \textit{Occupational income inequality} refers to economical distributions specific to the different occupations within population. It is measured with the network density of an income domination network.

% 6) - EDOIF overview of framework from input to output.
To achieve the income domination network, we present the usage of \textit{Empirical Distribution Ordering Inference Framework} (EDOIF). EDOIF is a framework that enables us to revisit income inequality from an occupation-based perspective. 

It takes a set of category-real ordered pairs, income and occupation. Then, performs a bootstrap for inferring confidence interval approximation of incomes by occupations. Consequently, the calculated confidence intervals infer dominant-distributions among the occupations. 

An income of occupation A is said to dominate an income of occupation B if there is a high probability that an income from the income distribution of A is greater than the expectation of income distribution of B but not vice versa (See Section \ref{orderingInference}). The resulting inference is a network of income dominant-distribution of occupations, income confidence intervals of occupations, income-mean-difference confidence intervals of occupation pairs, and occupation income orders \citep{EDOIF}.

% 7) - beyond output of EDOIF -> network density
The framework can also be applied in the inference of the degree of occupational income inequality in terms of income domination network. This is represented by the \textit{network density} which is a value representing a ratio between the number of occupation pairs that share an income dominating-dominated relationship over the total possible number of pairs. The network density discloses information about occupational income inequality in addition to the general income inequality as identified by the Gini coefficient. In a case where the Gini coefficient of a population is low, yet the network density is high, the population is said to have low-degree general income inequality but high-degree occupational income inequality. Low-degree general income inequality signifies that the overall income distribution is relatively proportional. High-degree occupational income inequality signifies that the income inequality is highly segregated by occupation. The data used in this study was obtained from Thailand household-population surveys from the Thai government in 2019~\cite{amornbunchornvej2021identifying}. 

% 8) - Contribution of this paper (drawback of gini -> EDOIF -> scatter as new analytical framework)
This paper offers new insight into evaluating the state of income inequality of a population by order of category. While the Gini coefficient can assess an entire population's state of inequality, it cannot distinguish minority groups of a certain category within the population. To do this, we use EDOIF to determine the domination order of occupations and its degrees of domination. Figure \ref{fig:pearson_correlation} presents a new analysis that combines the Gini coefficient with a measure of occupational income inequality. In the next section, the details of methodology are provided.

\section{Methodology}
\label{sec:methodology}

\subsection{Categorization of Occupations}
\label{OccuCategory}
The data provided for this study is the income of several individuals who are the head of their households along with their occupations grouped by the provinces of Thailand with the exception of Bangkok. The overview of this data can be found in the appendix.

The careers have been categorized as students (Student), animal farmer (AG-AnimalFarmer), plant farmer (AG-Farmer), fishery (AG-Fishery), orchardist (AG-Orchardist), peasant (AG-Peasant), business owner (Business-Owner), company employee (EM-ComEmployee), company officer (EM-ComOfficer), officer (EM-Officer), freelance (Freelance), merchant (Merchant), others (Others), and unemployment (Unemployment). Wherein the careers that are associated with the agricultural industry include animal farmer, farmer, fishery, orchardist, and peasant.

\subsection{Agricultural Categorization of Provinces} \label{AGCategory}
As the agricultural sector has been referred to as the engine of growth for Thailand \citep{AgricultureinThailand}, we find it appropriate to categorize the provinces of Thailand into agricultural (AG), mixed-agricultural (mixAG), and non-agriculture (nonAG) provinces. The categories are based on the ratio of head of households. A population over 66\% in either AG or nonAG occupations will be categorized accordingly. Provinces where the population of both agriculture and non-agriculture occupations does not exceed 66\% are referred to as mixed-agricultural provinces.

\subsection{Dominant-Distribution Ordering Inference} \label{orderingInference}
The Empirical Distribution Ordering Inference Framework (EDOIF) applies Dominant-Distribution Ordering Inference. It is the inference of an order of domination among several categories \(C\) by utilizing the Mann-Whitney Upper-tail Test with $\alpha=0.05$ to determine whether the distribution of category \(p\), \(D_p\), dominates that of category \(q\), \(D_q\), such that \(D_p \succeq D_q\). In our case, we use Dominant-Distribution Ordering Inference to determine a network of income domination among occupations of each province in Thailand under the null hypothesis that the income  of $p$ is less than or equal to the income of $q$ and, naturally, an alternate hypothesis that the the income  of $p$ is greater than the income of $q$ \citep{EDOIF}. The resulting dominant/non-dominant pairs are then summarized as income dominant-distribution networks for each province. Note that we use a domination network as a short term for an income dominant-distribution network. The network density in this work refer to the network density of an income dominant-distribution network, which has a value between 0 and 1. Higher network density implies higher number of occupational domination pairs. Hence, we use the network density as a measure for  occupational income inequality.  

\subsection{Aggregation Support Network} \label{supportNetwork}
Given the provincial domination networks as well as agricultural categorization of AG, mixAG, and nonAG provinces, we can now infer the Aggregation Support Networks. By applying the concept of Support in Association Rule Mining~\cite{Zhang2002}, the result is one network per agricultural categorization.

Let $T=\{t_i\}$ be a set of transactions  of provinces, which are provinces either in a specific category (i.e. AG, mixAG, or nonAG) or all provinces, in Thailand excluding Bangkok where $t_i$ is a transaction of ith province. Given $X,Y$ are occupations, within each $t_i=\{e(X,Y)\}$, $e(X,Y)\in t_i$ if an income of occupation $X$ dominates an income of occupation $Y$. The support is defined as a ratio of a number of times that an occupation pair $e(X,Y)$ appears in any $t_i\in T$ divide by a number of provinces. 

\begin{equation}
    supp(e(X,Y)) = \frac{ |\{t_i| t_i\in T \& e(X,Y) \in t_i\}| }{ |T|}
\end{equation}

Where $|\{t_i| t_i\in T \& e(X,Y) \in t_i\}|$ is a number of provinces that have an income of occupation $X$ dominates income of occupation $Y$, and $|T|$ is a number of provinces. With the support value of each domination pair and a support threshold of 0.5, the aggregation support networks can now be concluded for each agricultural categorization. In the aggregation support network, the nodes are occupations and there is an edge $\hat{e}(X,Y)$ in the network if the support $supp(e(X,Y))$ is greater than 0.5.

\section{Results}
\label{sec:results}

% 1) Exploratory Data Analysis ->  visualizations and summary statistics
Through the usage of Exploratory Data Analysis, the process of investigating data for patterns \citep{EDA}, we have created visual representations and summary statistics to represent our findings with various forms of analyses. 

This study compiles maps of Thailand displaying the locations of general income inequality and occupational income inequality among all provinces in Thailand with the exception of the capital city of Bangkok. Additionally, the distribution of Gini coefficients, network density, and average income rate is also calculated. Finally, among these key socioeconomic indicators, the correlations distinguished by agricultural provinces for Thailand in 2019 is presented. In this work, the provinces are grouped into three types based on the majority of occupations in provinces as explained in Section \ref{AGCategory}. The categories are agricultural provinces (AG), non-agricultural provinces (nonAG), and mixed provinces (mixAG).

\begin{figure}
    \centering
    \includegraphics[width=\textwidth,height=\textheight,keepaspectratio]{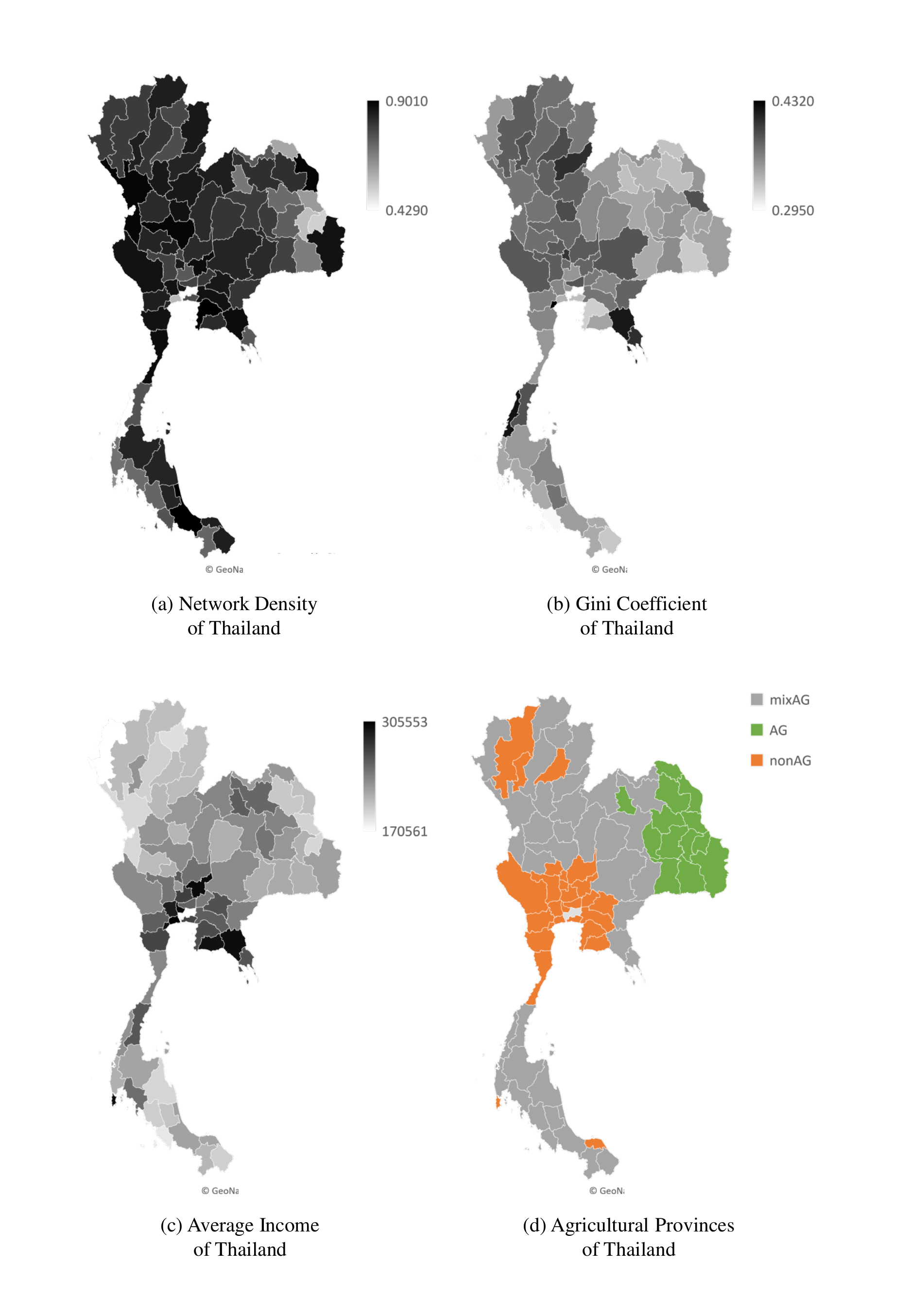}
    \caption{Maps of Thailand depicting (a) Network Density, (b) Gini Coefficient, (c) Average Income, and (d) Agricultural Provinces}
    \label{fig:maps_of_thailand}
\end{figure}

% 2) Map of Thailand -> AG prov in NE, nonAG prov in Central + NW
The maps of Thailand illustrate location of provinces' Gini coefficients, average incomes, categorization of agricultural provinces, and network densities as shown in Figure \ref{fig:maps_of_thailand}.

In the aspect of occupational income inequality (Figure~\ref{fig:maps_of_thailand}a), the areas that have high network densities are distributed across the country. For the general income inequality (Figure~\ref{fig:maps_of_thailand}b), however, the high-Gini-coefficient areas are around the upper-central provinces, the east, and the north of country.
In Figure~\ref{fig:maps_of_thailand}c, the areas with high average income are around the central of country. For agricultural types of provinces (Figure~\ref{fig:maps_of_thailand}d), the majority of agricultural provinces are located in Northeastern Thailand whereas non-agricultural provinces are generally clustered near the capital city, in Central Thailand, and Northwestern Thailand. 

Briefly, the high-Gini-coefficient and high-average-income provinces are distributed towards non-agricultural provinces, while high network-density provinces are everywhere. Figures \ref{fig:maps_of_thailand}a and \ref{fig:maps_of_thailand}b provide a clear view on discrepancy between network density and the Gini coefficient. For instance, Chonburi is ranked the top 5\% highest network density and bottom 5\% lowest Gini coefficient whereas Ranong is ranked the bottom 5\% lowest network density and top 5\% highest Gini coefficient. This quantitatively implies that the network density offers insight beyond the Gini coefficient.

Since areas categorized by these four aspects\textemdash network density, Gini coefficients, average income, and agricultural categories\textemdash that do not overlap completely, they play a different role in an econometric explanation of each area.

\begin{figure}
    \centering
    \includegraphics[width=\textwidth]{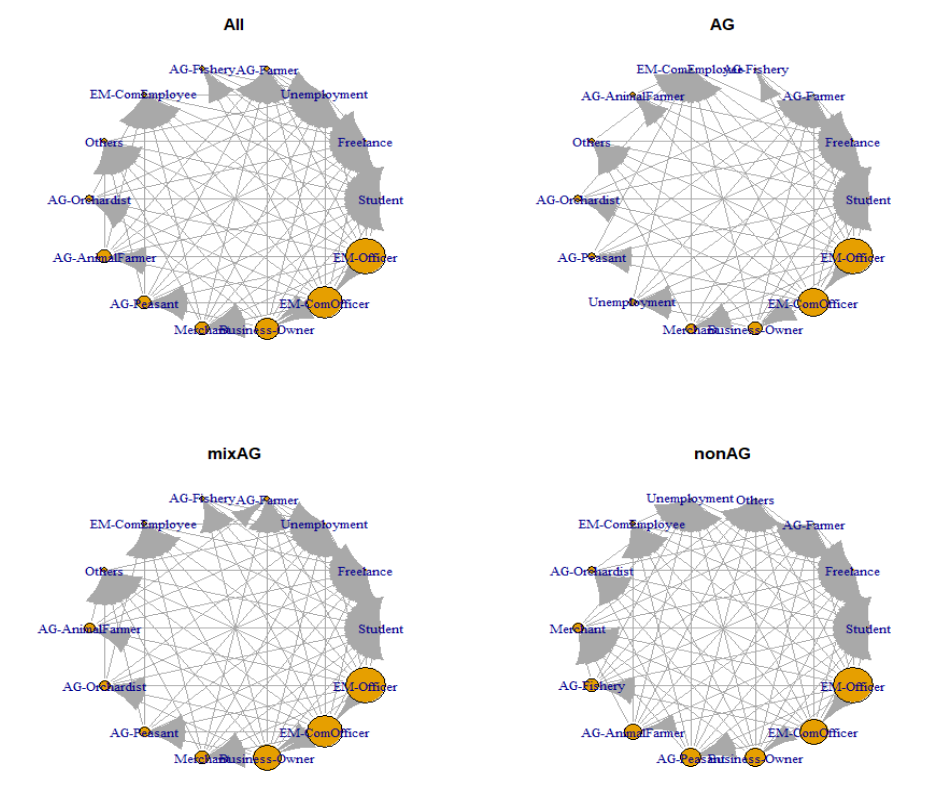}
    \caption{Aggregate Support Networks for four types of provinces: all provinces (All), agricultural provinces (AG), non-agricultural provinces (nonAG), mixed provinces (mixAG)}
    \label{fig:agg_supp_net}
\end{figure}

% 3) Histogram of gini, netDen, ave income -> netDen is high throughout Thailand, ave income "normal" distributed and high at central TH, gini "normal" distributed and high at not AG.
In Figure \ref{fig:agg_supp_net}, the domination networks are presented (See Section \ref{supportNetwork} for derivative). There is no significant difference among the network structure of agricultural types of provinces. This implies that the common structure of occupational inequality exist for the entire country (i.e. EM officer occupation dominates any occupation and the student occupation is also dominated by any occupation in any province).  

\begin{figure}
    \centering
    \includegraphics[width=\textwidth]{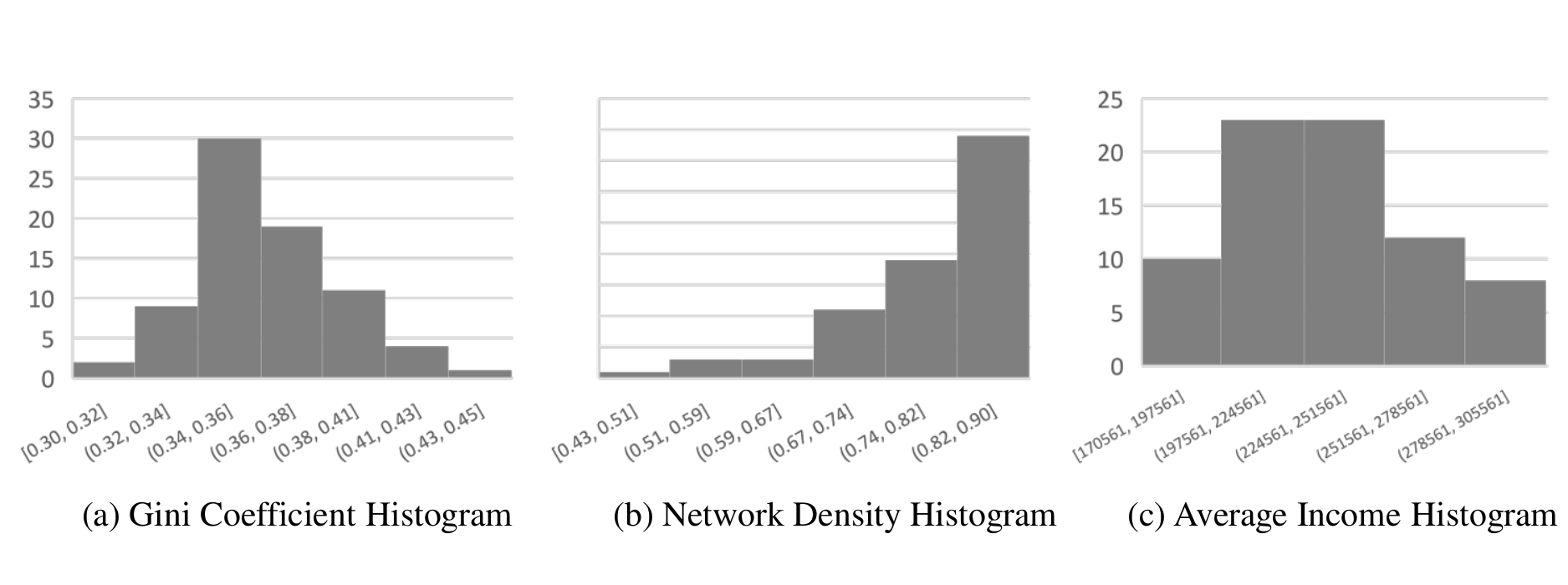}
    \caption{Frequency distribution of (a) Gini Coefficient, (b) Network Density, and (c) Average Income}
    \label{fig:histograms}
\end{figure}

Figure \ref{fig:histograms} represents histograms of Gini coefficients, network density, and average incomes among provinces. In general, the network densities are noticeably high and between 0.78 and 0.85, the top 15\% of the range, points to a feasibility that income inequality specific to distribution by occupation is a major concern as can be seen from Figure \ref{fig:histograms}b. On the other hand, Figures \ref{fig:histograms}a and \ref{fig:histograms}c display near normal distributions of Gini coefficients and average income histograms.

\begin{figure}
    \centering
    \includegraphics[width=\textwidth]{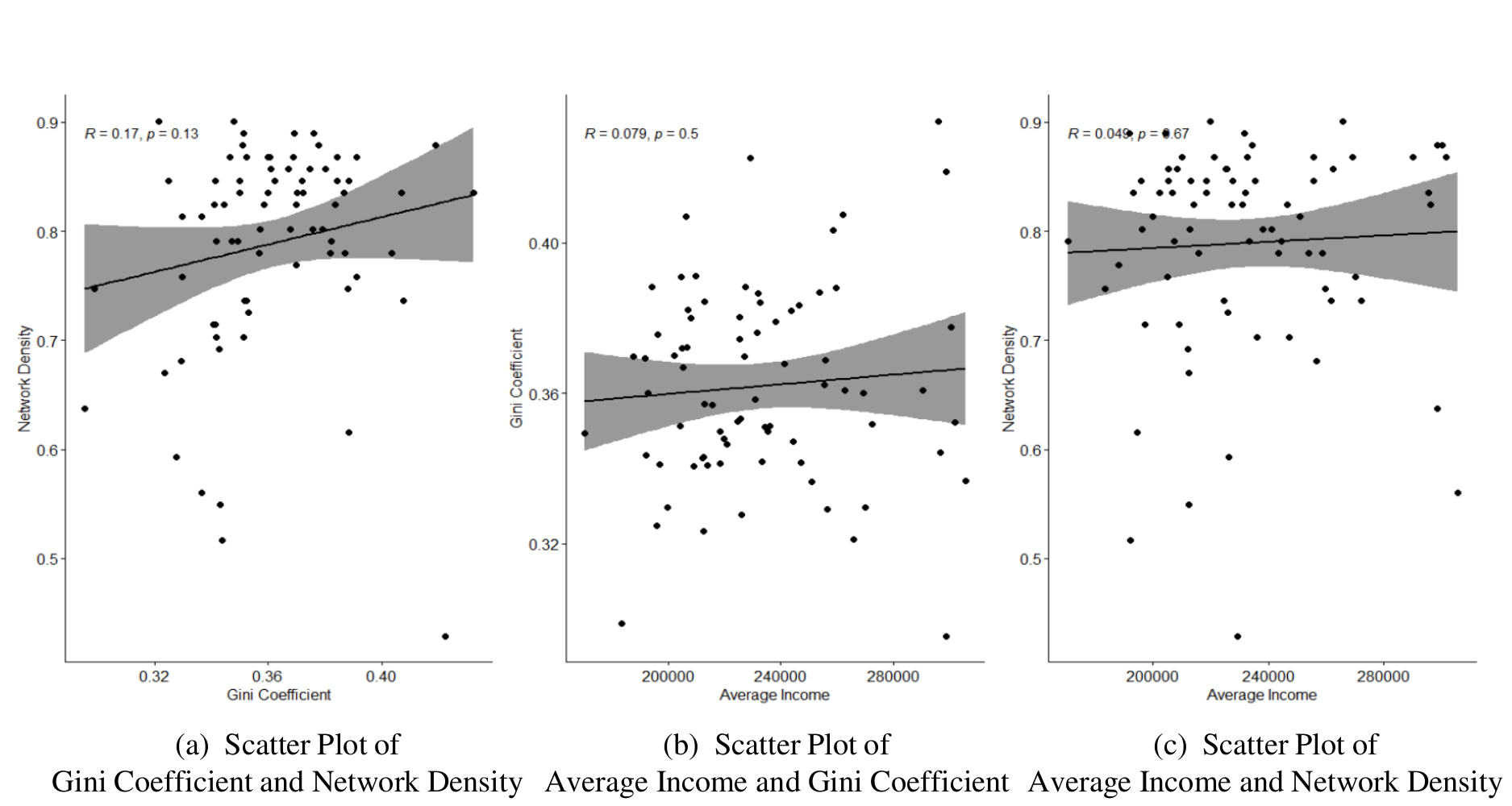}
    \caption{Visual inspection of correlation between (a) Gini Coefficient and Network Density, (b) Average Income and Gini Coefficient, and (c) Average Income and Network Density}
    \label{fig:pearson_correlation}
\end{figure}

% 4) Pearson correlation -> gini and netDen not related -> explore significance of netDen
In the aspect of association among network density, Gini coefficient, and incomes, the results are shown in Figure \ref{fig:pearson_correlation}.  The Pearson correlation between the network density and Gini coefficient is 0.17, the correlation between the Gini coefficient and average income is 0.079, and finally, the correlation between network density and average income is 0.049. 

According to the work in \citet{Sullivan2012}, the effect sizes of these \emph{r} values below 0.2 indicate small effect sizes. Thus, the measure of income inequality specific to occupation does not appear to be statistically related to the current measures of general income inequality known as the Gini coefficient and can therefore infer information about a population that is previously unknown. To explore the significance of occupational income inequality in comparison to general income inequality, we performed further analysis between the network density and Gini coefficient w.r.t. the percentage of population in the agricultural industry.

\begin{figure}
    \centering
    \includegraphics[width=\textwidth]{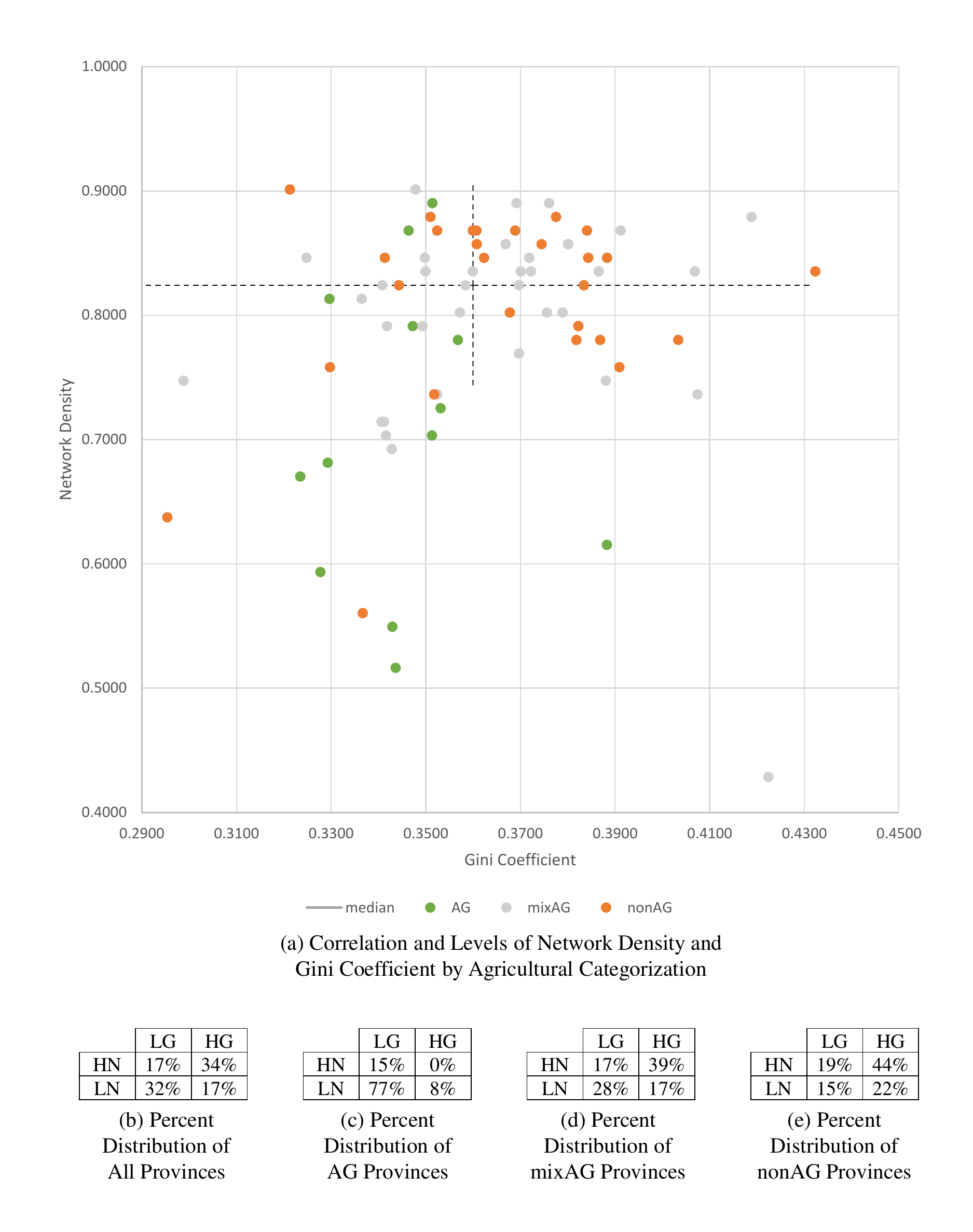}
    \caption{Relationship of Occupational Income Inequality and General Income Inequality in Relation to Agricultural Categories}
    \label{fig:netDen_gini_LH}
\end{figure}

% 5) gini vs netDen plot -> LGHN exists, AG likely LGLN and nonAG likely HG
In Figure \ref{fig:netDen_gini_LH}a, a plot of the network density and Gini coefficient distinguishing provinces are colored by agricultural categories. It reveals a convincing pattern between agricultural categorization and general income inequality in addition to agricultural categorization and occupational income inequality.

The figure illustrates the provincial network density w.r.t. Gini Coefficient where green indicates agricultural (AG) provinces and orange is indicative of non-agricultural (nonAG) provinces. The grey data points indicate mixed agricultural and non-agricultural (mixAG) provinces. The line which differentiates high as opposed to low for either axes marks the median. According to Figure \ref{fig:netDen_gini_LH}b, 17\% of all provinces in Thailand identify as populations of low Gini coefficients (LG), and thus low-level general income inequality, yet high network density (HN), and therefore high-level occupational income inequality. These are provinces that would be previously indiscernible to policy makers by analyzing Gini coefficient alone. Additionally, from Figure \ref{fig:netDen_gini_LH}c, as high as 92\% of all AG provinces appear to exhibit LG and among these provinces, 77\% also exhibit low network densities (LN) implying low-level occupational income inequality which suggests that AG provinces in Thailand likely exude LGLN cases. As for nonAG provinces (Figure \ref{fig:netDen_gini_LH}e), 66\% show high Gini coefficients (HG) thus high-level general income inequality. 

% LGHN - formerly "invisible" province
In accordance to current measures, low-level general income inequality suggests that the population under observation is unaffected by inequality of income. However, when the same population exhibits high network densities, this points to an instance where the population seemingly has low-level inequality yet is significantly influenced by differences in occupational inequality. This is seen in occurrence among the 17\% of all provinces in Thailand (Figures \ref{fig:netDen_gini_LH}b), which are highly troubled with unequal wage dispersion by occupation yet undetectable by general income inequality. For instance, Songkhla province is a LGHN province.

\begin{figure}
    \centering
    \includegraphics[width=\textwidth]{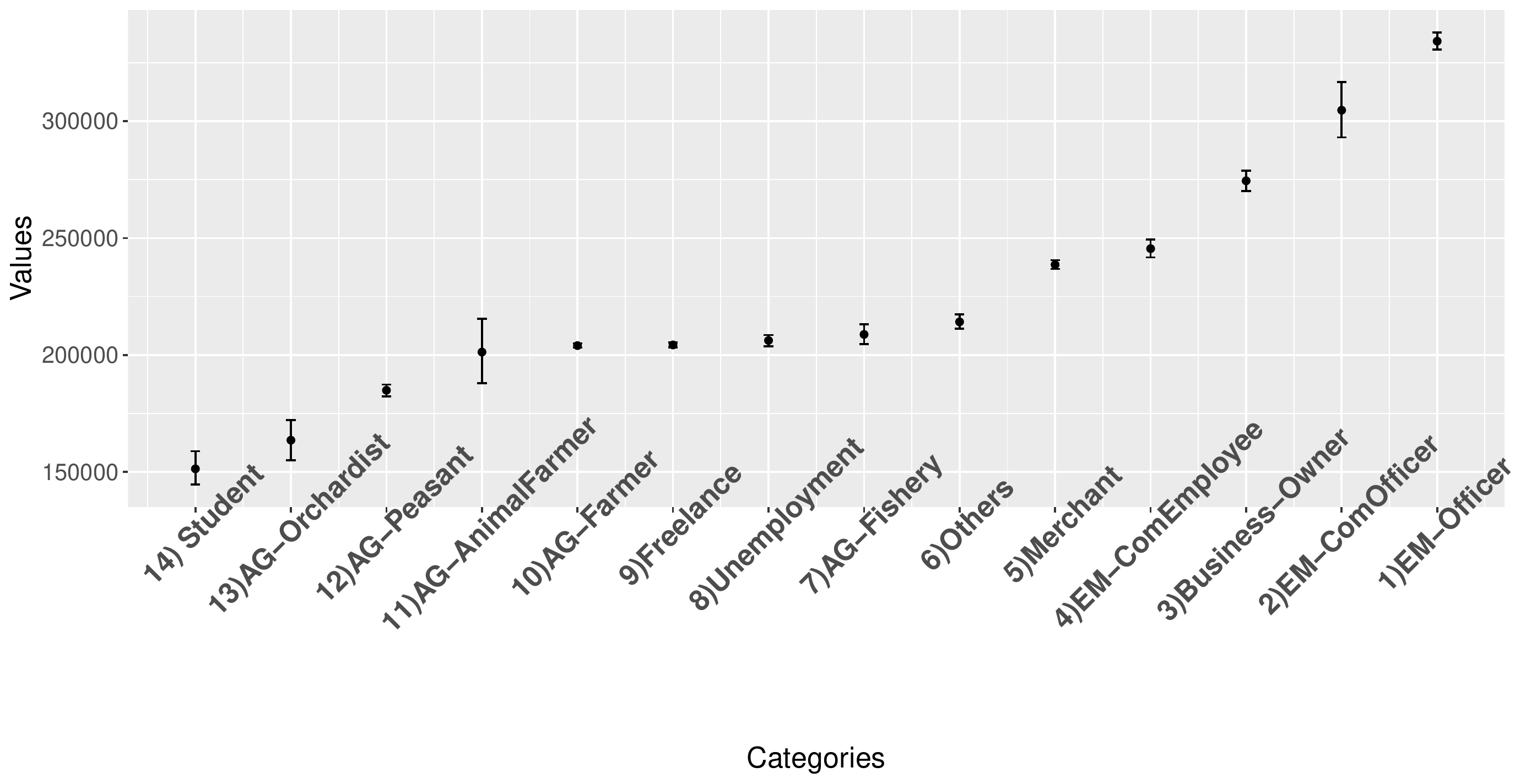}
    \caption{Income Confidence Interval of occupations in Songkhla Province}
    \label{fig:LGHN_CI_occu}
\end{figure}

\begin{figure}
    \centering
    \includegraphics[width=\textwidth]{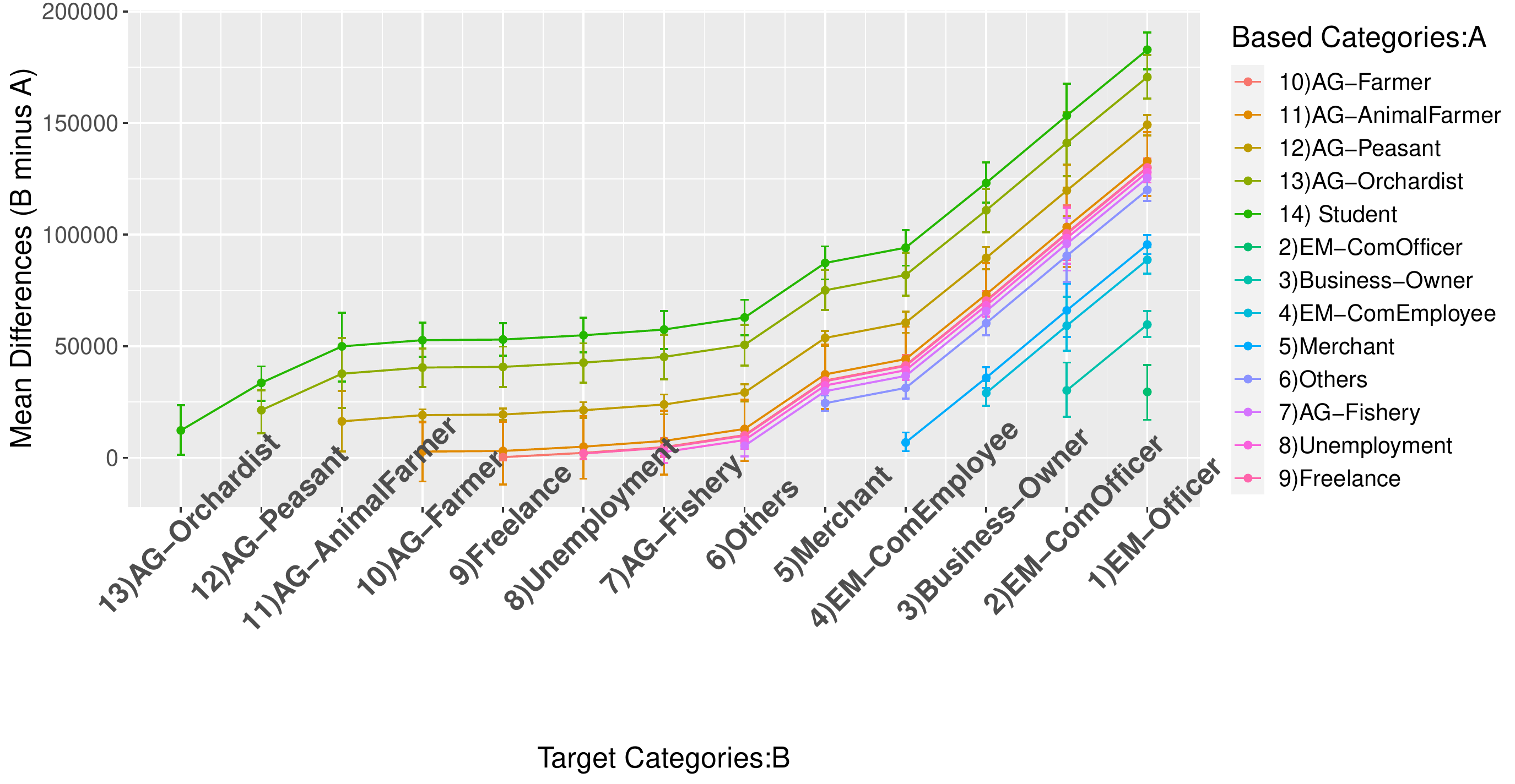}
    \caption{Mean-difference Income Confidence Interval of occupation pairs in Songkhla Province}
    \label{fig:LGHN_MDCI_occuPairs}
\end{figure}

% example case that have low gini but high net den. 
Songkhla is a mixed-agricultural (mixAG) province (See percent distribution in Figure \ref{fig:netDen_gini_LH}d). The number of occupations are relatively equally distributed between agricultural and non-agricultural occupations. The low provincial Gini coefficient of 0.3479 shows a relatively equal distribution of income across the entire population. Figure \ref{fig:LGHN_CI_occu} illustrates occupation ordering as a result of approximately 280,000 head of households in the province in 2019. The income dominant occupation is "EM-Officer" which dominates all other occupations with an overall income above 325,000 THB or roughly 9,000 USD. The high provincial network density of 0.901 demonstrates that income between occupations of Songkhla are highly unequally distributed. Moreover, the magnitudes of mean-difference income confidence interval of all pairs of occupations in Songkhla from Figure \ref{fig:LGHN_MDCI_occuPairs} portrays that all the occupation pairs have incomes gaps of at least 25,000 THB or about 750 USD.

%HGLN
% On the other hand, as summarized by Figure \ref{fig:levels_of_inequality}, high Gini coefficients suggest high-level general income inequality but when these populations also show low-level occupational income inequality with low network densities, the population of interest is significantly affected by income inequality generally, but segregation of occupation is not outstanding among the reasons why. For example, 22\% of all nonAG provinces in Thailand have high-level general income inequality but do not show high-level occupational income inequality.

% \begin{figure}
%     \centering
%     \includegraphics[width=\textwidth]{figures/levels_of_inequality.JPG}
%     \caption{Summary of Implications on Levels of Inequality}
%     \label{fig:levels_of_inequality}
% \end{figure}

% LGLN - example of AG provinces that are likely this case
On the other hand, demonstrating a province that is not significantly affected by income inequality generally nor by occupation-specific means, is where low Gini coefficients suggest low-level general income inequality and when these populations also show low-level occupational income inequality with low network densities, the population of interest is under relative ideal income conditions. For example, Amnat Charoen province is among one of the 77\% of AG provinces in Thailand that have low-level general income inequality and also show low-level occupational income inequality.

\begin{figure}
    \centering
    \includegraphics[width=\textwidth]{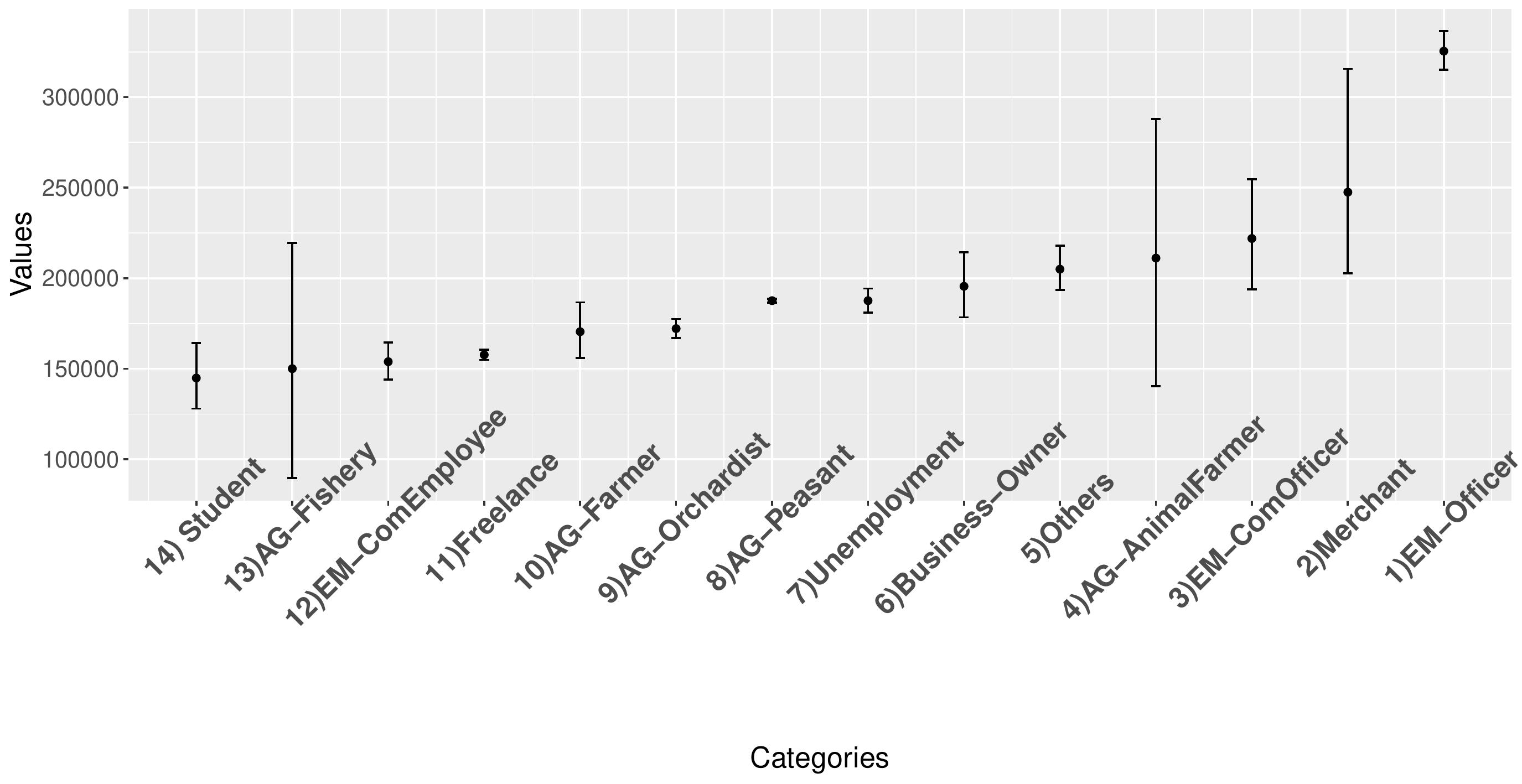}
    \caption{Income Confidence Interval of occupations in Amnat Charoen Province}
    \label{fig:LGLN_CI_occu}
\end{figure}

\begin{figure}
    \centering
    \includegraphics[width=\textwidth]{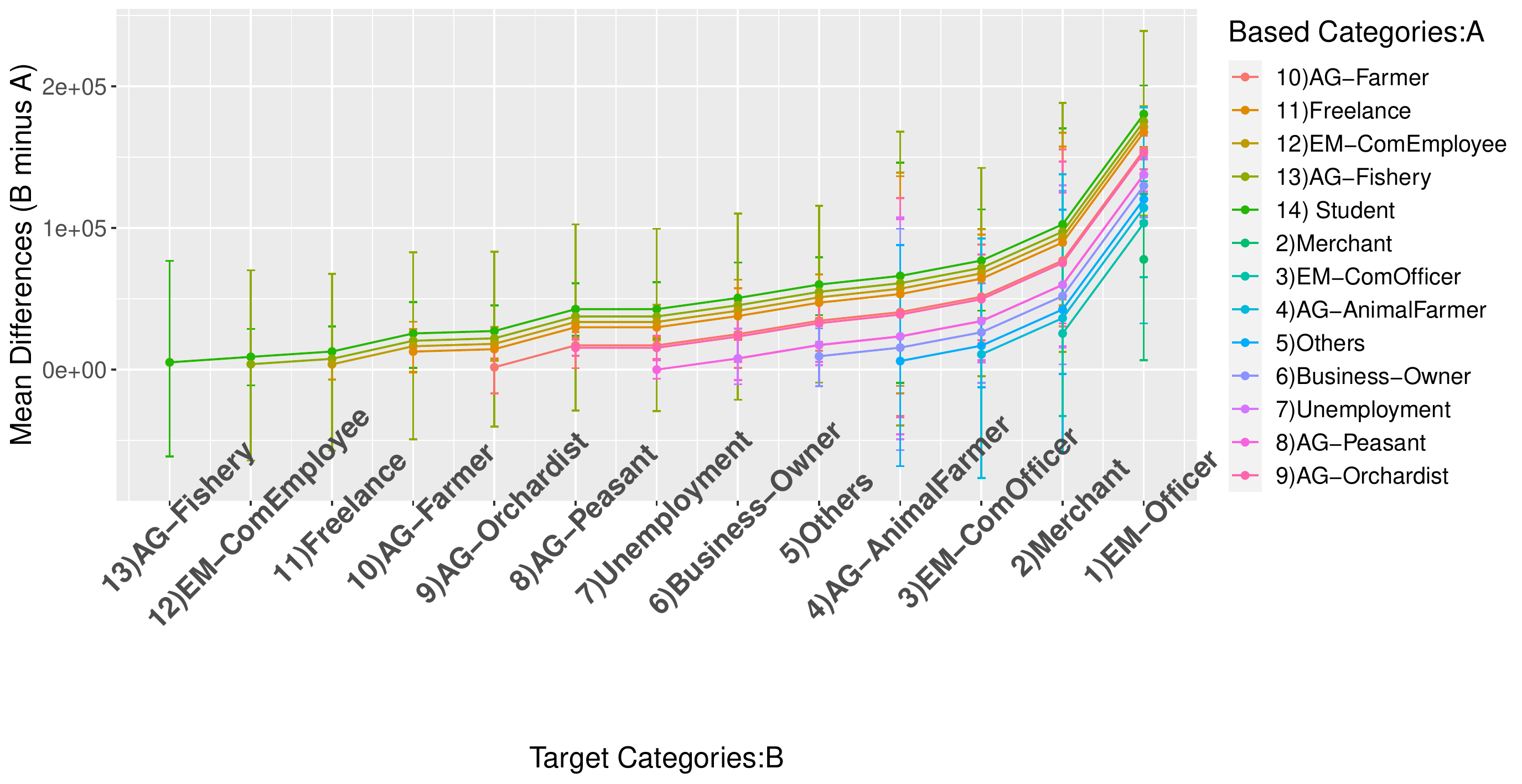}
    \caption{Mean-difference Income Confidence Interval of occupation pairs in Amnat Charoen Province}
    \label{fig:LGLN_MDCI_occuPairs}
\end{figure}

Amnat Charoen is an AG province whose mild gaps of occupational income inequality is illustrated in Figures \ref{fig:LGLN_CI_occu} and \ref{fig:LGLN_MDCI_occuPairs}. With a low Gini coefficient of 0.3437, the income distribution of the province is fairly equally distributed. The occupation ordering shows that the dominant occupation is "EM-Officer" with an overall income above 300,000 THB or roughly 8,900 USD. The low provincial network density of 0.517 suggests that income between occupations are relatively equally distributed throughout Amnat Charoen. The magnitudes of mean-difference income confidence interval of most pairs of occupations portrays that income gaps of all occupation pairs are at least 50,000 THB or about 1500 USD. Comparing against Songkhla Province, Amnat Charoen has less number of income inequality of occupation pairs but when it has a gap, it has larger gaps of income.

% chi-square supports that AG likely LGLN and nonAG likely HG
Given that the null hypothesis states the agricultural categories of provinces is independent to the low-high categories of the Gini coefficient and network density, the alternative hypothesis states that the two are dependent. A Chi-Square Independence Test with the significant level $\alpha=0.05$ supports that the categorizations of AG provinces and their corresponding low-high Gini coefficient and network density are not independent of each other. This supports the assumptions from our above statements that AG provinces are associated with low general income inequality and nonAG provinces are associated with high general income inequality yet for occupational income inequality only AG provinces show significant favor towards low occupational income inequality.

\section{Discussion}
\label{sec:discussion}

% 1) LGHN supports importance of inter-occupational inequality. Most AG prov are LGLN and show relative ideal living conditions in terms of income inequality.
The findings in this study distinguish the provinces of Thailand into groups characterized by levels of general and occupational income inequality. A combination of two groups show results that suggest an ideal condition for lowering overall income inequality in Thailand and perhaps other countries with similar measurement index results. When general income inequality is unsubstantial and occupational income inequality is serious, such as the case of Songkhla shown in Figures \ref{fig:LGHN_CI_occu} and \ref{fig:LGHN_MDCI_occuPairs}, the condition of income inequality is misleading to residents and policy makers. This supports the statement that inter-occupational inequality is a significant aspect of income inequality that should not be overlooked \citep{kambourov2009}. On the other hand, when both the general and occupational income inequality is low as is in Figures \ref{fig:LGLN_MDCI_occuPairs}, this points to a province with ideal living conditions where rates of unemployment is comparatively lower and overall income inequality is low-level.

% 2) H-T model explains effects of AG-to-nonAG (rural-to-urban) migration
The Harris-Todaro Migration Model is a predictive model that explains the rural-to-urban migration behavior of a population in terms of agricultural and non-agricultural wages and availability. It is proposed that the rate at which residents will migrate from rural to urban locations is dependent on an individual's perceived value of migration. This includes information availability heuristics, expected returns from migration, and expected cost of migration leading up to a migration decision. However, this analysis does not take into account the difficulty for a typical migrant to secure a full-time employment upon migration nor the consequences of unemployment while seeking a job. These factors lead to an issue of unemployment. As a result, the decision for rural-to-urban migration made by an individual is not necessarily beneficial towards economic development of the society \citep{Todaro2015}.

% 3) rural-urban migration still happens in Thailand despite harmful consequences to economic development -> suggest policy to discourage migration
The National Migration Survey of Thailand 2019 shows evidence in compliance with the Harris-Todaro Migration Model. The amount of people moving into Thailand's central region makes up approximately 45\% of all migrants where Bangkok alone accounts for 19\% of these migrants. The second largest influx of incoming migrants is located in Northeastern Thailand at 26\% \citep{MigrationSurvey}. Both of which are locations of non-agricultural provinces. From our results, these nonAG provinces are likely to display communities of high general income inequality. With the current living conditions from a resident viewpoint, rural-to-urban migration is a probable decision made by the individual living in Thailand. Subsequently, the over-saturation of rural-to-urban migrants will continue to contribute towards rural-urban segregation of wage dispersion and overall income inequality of the population \citep{reda2012}. Countermeasures suggested by \citet{Todaro2015} include discouraging rural-to-urban migration by directing governmental policy to fight poverty towards the agricultural or non-agricultural sector of the areas specifically where the majority of the poor reside. 

% AG most likely has LGLN -> suggest policy target at LGLN areas
In this study, AG provinces have low Gini coefficient and low occupational domination network density, which suggests that AG provinces experience low general and occupational income inequality. It is established that income inequality is associated with severe economic consequences. Populations with higher income inequality grow slower than those of lower income inequality \citep{buttrick2017}. Reduced inequalities are less sustainable in high-inequality populations \citep{ostry2011}. An increase in GDP frequently concentrates towards a small percentage of the highly unequal population \citep{piketty2014}. Lastly, poor health is also very common among populations of high inequality \citep{pickett2014}. This remains true for both low-income and high-income individuals within a highly unequal population relative to populations of lower income inequality \citep{subramanian2006}. Therefore, areas of low general and occupational income inequality (i.e. AG provinces) may be a suitable area for policy makers to encourage residents remain. This is consistent with the Harris-Todaro Migration Model. 

% 4) how information contributes to policy makers
From the perspective of the policy makers, the Gini coefficient cannot indicate where the majority lies in terms of general income inequality but the usage of EDOIF to analyze occupational income inequality allows us to see minority groups by order of income oppression. Among the majority of income confidence intervals of all provinces, there exists a global structure that  remains intact throughout. "Student" and "EM-Officer", for the most part, rank most and least dominated, respectively throughout Thailand. By acknowledging both measurement indices, policy makers can utilize this information to develop policies that leaves no minority behind.

\section{Conclusion}
\label{sec:conclusion}

% How this study contributes -> results -> discussion on problem -> solution from our best knowledge
This study investigated how \textit{occupational income inequality} provides insight into \textit{general income inequality} as measured by the Gini coefficient. Our study showed that most agricultural provinces display ideal levels of both occupational and general equality in comparison to non-agricultural provinces which tend to experience a strong indication of general income inequality. Moreover, some non-agricultural provinces show the degree of high occupational income inequality even if they have low general income inequality. Despite these conditions, the Thai population still frequently choose to migrate towards the locations of growing income inequality for individual gain. Thus, occupational income inequality can be utilized to assist policy makers in decision-making for economic development. From our best knowledge, a solution may be to align the benefits of individuals to those of the national economy. This might be by supporting minority occupations in areas of massive inequality. For example, students, who face income inequality issue, residing in areas of high occupational income inequality might get more support. Policy makers can also discourage rural-to-urban migration by distributing the benefits of rural income to a wider audience so that the option naturally becomes more alluring to individuals.

\bibliographystyle{elsarticle-num-names} 
\begin{CJK}{UTF8}{min}

%\bibliography{cas-refs}
\end{CJK}

\end{document}